\documentclass{iau}

\usepackage{amsmath}
\usepackage{graphicx}
\usepackage{multirow}

\usepackage{caption,subcaption}

\begin{document}

\lefttitle{Natalie Myers, et al.}
\righttitle{Optical Neutron Capture Abundances of Open Cluster Stars}

\jnlPage{1}{7}
\jnlDoiYr{2025}
\doival{10.1017/xxxxx}
\volno{395}
\pubYr{2025}
\journaltitle{Stellar populations in the Milky Way and beyond}

\aopheadtitle{Proceedings of the IAU Symposium}
\editors{J. Mel\'endez,  C. Chiappini, R. Schiavon \& M. Trevisan, eds.}

\title{Beyond OCCAM: Measuring Optical Neutron Capture Abundances of Open Cluster Stars}

\author{Natalie Myers$^1$, Sarah Loebman$^2$, Henrique Reggiani$^3$, Peter Frinchaboy$^{1,5}$}
\affiliation{$^1$Department of Physics \& Astronomy, Texas Christian University, Fort Worth, TX 76129, USA}
\affiliation{$^2$Department of Physics, University of California, Merced, 5200 Lake Road, Merced, CA 95343, USA}
\affiliation{$^3$Gemini Observatory/NSF’s NOIRLab, Casilla 603, La Serena, Chile}
\affiliation{$^4$The Observatories of the Carnegie Institution for Science,  Pasadena, CA 91101, USA}
\affiliation{$^5$Canada-France-Hawaii Telescope, 65-1238 Mamalahoa Highway, Kamuela, HI 96743, USA}

\begin{abstract}
Open clusters have long been used to determine ages of stars, as well as calibrate stellar evolution models and other methods of age-dating stellar groups, e.g., gyrochronology, asteroseismology, and chemical clocks.  In this work, we have obtained new high-resolution ($R \ge 50,000$), high-$S/N$, optical data for 3+ stellar members in open clusters, using Keck/HIRES, with membership derived from the Open Cluster Chemical Abundances and Mapping (OCCAM) survey.  From these new Keck/HIRES data, we have derived neutron capture abundances for stars in seven distant outer Galaxy open clusters.
\end{abstract}

\begin{keywords}
stars: abundances,
Galaxy: evolution,
open clusters and associations: general
\end{keywords}

\maketitle
% \vskip -0.1in

\section{Introduction}

The chemistry of stars in the Milky Way is a powerful tool for exploring the enrichment history of the Galaxy.  With the all-sky spectroscopic surveys that are currently available to us, using chemistry as a means to identify the ages of field stars is becoming more feasible.  However, in order to build an accurate and reliable age-chemistry relationship (or chemical clock), it first needs to be calibrated with reliable age-tracers.

The goal of the Open Cluster Chemical Abundance and Mapping \citep[OCCAM;][]{OCCAM1} survey is to provide a high-quality catalog of open clusters to study the chemical enrichment history of the Milky Way. OCCAM used data from SDSS/APOGEE DR17 \citep{DR17_APOGEE} and the 5D kinematic properties from Gaia EDR3 \citep{Gaia} to determine cluster membership used in this work. 
APOGEE provides abundances for numerous alpha ([O, Mg, Si, S, Ca, Ti]), iron-peak ([Ni, Co, Mn, Cr, V]), odd-z ([K, Al, Na]), and light elements ([C, N]).  OCCAM has already been used to measure light element abundances and measure the Galactic abundance gradient \citep[OCCAM-VI;][]{OCCAM_Myers} and to calibrate the [C/N] chemical clock \citep[OCCAM-VII;][]{OCCAM_Spoo} using the APOGEE data.  However, abundances for heavier elements are less reliable and available in the infrared. 
 
In this work, we acquire $r-$ and $s-$ process elements for open cluster members, to build a reliable, age-datable sample of all major element groups.  Not only can this sample be used to explore the chemical enrichment history of the Milky Way, but it can also be used to characterize age-chemistry relationships (chemical clocks).  Adding the neutron capture element group may also help distinguish otherwise chemically similar stars \citep[e.g.,][]{Manea}.

%\clearpage
\section{Methods}
\begin{figure}
\centering
\includegraphics[scale=.23]{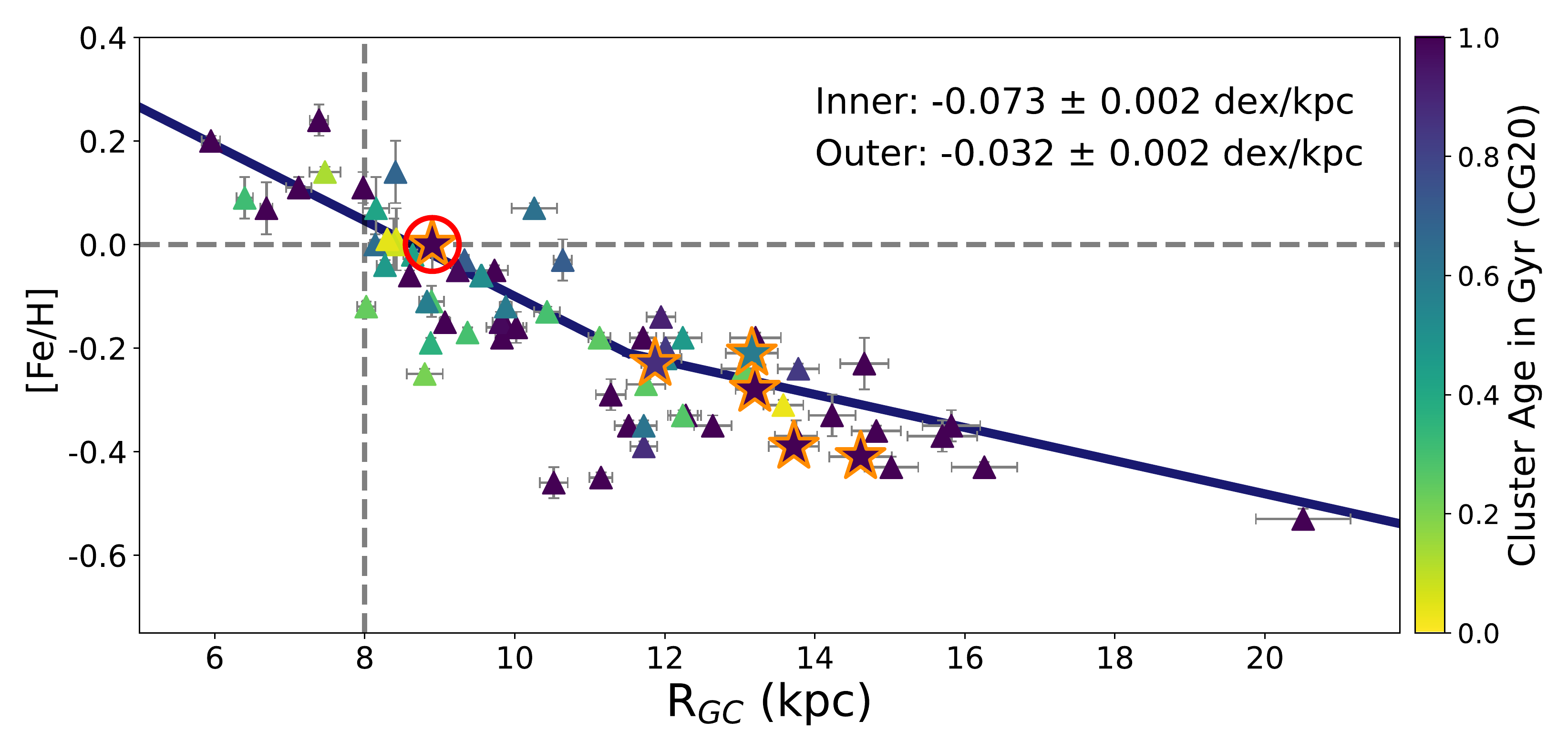}
\caption{The OCCAM metallicity gradient from \citet{OCCAM_Myers} with follow-up clusters identified. All OCCAM high-quality clusters are shown as colored triangles with color indicating cluster age, saturating at 1 Gyr (the maximum age in this sample is 9 Gyr). The clusters marked with stars represent those which we have re-observed with Keck HIRES in the optical regime, presented in this work. The circled cluster in the solar neighborhood is our calibrator, M67. The metallicity gradient fit is shown in blue with the slopes of the fit in the upper right. We specifically targeted old open clusters at large Galactic radii to extend the neutron capture trends.}
\label{gradient}
\end{figure}

We utilize the SDSS-IV/APOGEE-based OCCAM survey as the foundation for our optical follow-up observations (see Figure \ref{gradient} for the full sample). For each cluster, we identify 3+ high-quality stellar members, supplementing with Gaia-based \citet{CG20} high-quality cluster members where necessary.  Here, we focus on data taken using the Keck I telescope and HIRES spectrograph in Hawai’i, where we acquired high-resolution ($R\sim$ 50,000+) spectroscopic data with a signal-to-noise ratio of at least 75 at 5500\AA~for faint ($13 < V_{mag} < 15.5$) cluster members. We observed a total of 29 stars, covering 8 clusters, 16 of which are used in our preliminary analysis here (shown as stars in Figure \ref{gradient}). 

For the reductions of the data, we start with the standard HIRES MAKEE reductions for these stars. We then use IRAF and PyRAF to wavelength correct, median combine, normalize, and RV correct each star with the standard \textit{echelle} and \textit{rv} packages. 

The abundance analysis was conducted using the Brussels Automatic Code for Characterizing High-accUracy Spectra \citep[BACCHUS,][]{BACCHUS}. 
The BACCHUS code uses Turbospectrum \citep{Turbospec} and MARCS models \citep{Marcs} to measure abundances by spectral synthesis. For a given spectrum and initial guess (given by APOGEE stellar parameters), BACCHUS first determines the model that best fits the stellar parameters (e.g., T$_{\rm eff}$, log(g), and [Fe/H]), then measures abundances line-by-line for the desired elements.  For these preliminary results, we derive abundances for over 12 elements using the Gaia-ESO line list. 

\section{Results}

\begin{figure}[htb]
    \centering 
\includegraphics[width=\linewidth]{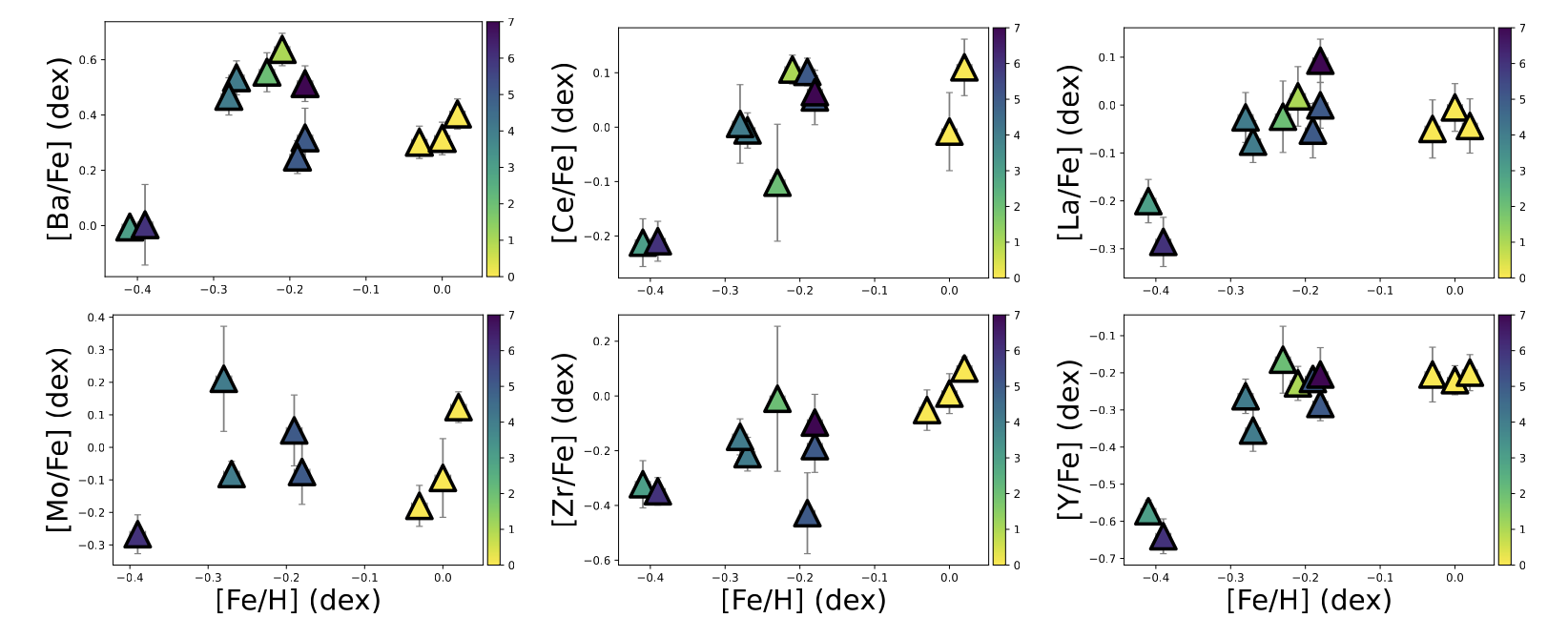}

\caption{The neutron-capture elements Ba, Ce, La, Mo, Zr, and Y for the stars of our eight clusters plotted with metallicity on the x-axis. The colors correspond to the cluster the stars belong to. In this analysis, not all stars have well-measured abundances for all elements. 
}
\label{abundances}
\end{figure}

From these data, we have measured abundances for the $s-$process elements Ce, Ba, Y, Mo, La, and Zr in seven distant open clusters. These abundances are presented in Figure \ref{abundances}, where different members of the same cluster have a matching color.  In general, we find a non-linear trend of increasing abundance with cluster metallicity ([Fe/H]). Although this trend is noticeable in all elements, it is shallower in molybdenum and zirconium. 

We will improve the precision of this analysis by updating to an improved line list and also derive $r-$process abundances, such as europium, that will be presented in Myers et al. {\em in prep}.

\clearpage
\section{Acknowledgments}

The authors extend our gratitude to Catherine Manea and Keith Hawkins for their invaluable guidance with BACCHUS, and to Matthew Shetrone for his assistance with IRAF.\\

We acknowledge funding from the National Science Foundation (AST-1715662 \& 2206541) as well as the Aspen Center for Physics PHY-1607611). Funding for the Sloan Digital Sky Survey IV has been provided by the Alfred P. Sloan Foundation, the U.S. Department of Energy Office of Science, and the Participating Institutions. SDSS acknowledges support and resources from the Center for High-Performance Computing at the University of Utah. The SDSS Web site is \href{http://www.sdss4.org/}{http://www.sdss4.org/}.

\end{document}